\documentclass[sigconf, authorversion, nonacm]{acmart}

\AtBeginDocument{%
  }

\setcopyright{acmlicensed}
\copyrightyear{2018}
\acmYear{2018}
\acmDOI{XXXXXXX.XXXXXXX}

\acmISBN{978-1-4503-XXXX-X/18/06}

\begin{document}

\title{Vsens Reality: Blending the Virtual Sensors into XR}


\author{Fengzhou Liang}
\affiliation{%
  \institution{Keio University}
  \city{Yokohama}
  \country{Japan}
  }
\email{liangfz@keio.jp}

\author{Tian Min}
\affiliation{%
  \institution{Keio University}
  \city{Yokohama}
  \country{Japan}
  }
\email{welkinmin@keio.jp}

\author{Yuta Sugiura}
\affiliation{%
  \institution{Keio University}
  \city{Yokohama}
  \country{Japan}
  }
\email{sugiura@keio.jp}

\renewcommand{\shortauthors}{Trovato et al.}

\begin{abstract}
In recent years, virtual sensing techniques have been extensively studied as a method of data collection in simulated virtual spaces for the development of human activity recognition (HAR) systems. To date, this technique has enabled the transformation between different modalities, significantly expanding datasets that are typically difficult to collect. However, there is limited research on how to make virtual sensors more easy-to-use or effective as tools for making sense of the sensor data. The context-awareness and intuitiveness of XR make it an ideal platform for virtual sensors. In this work, we demonstrate, \textit{Vsens Reality}, the use of virtual sensors under the XR context as an augmentation tool for the design of interactive systems.
\end{abstract}




\begin{teaserfigure}
  \includegraphics[width=\textwidth]{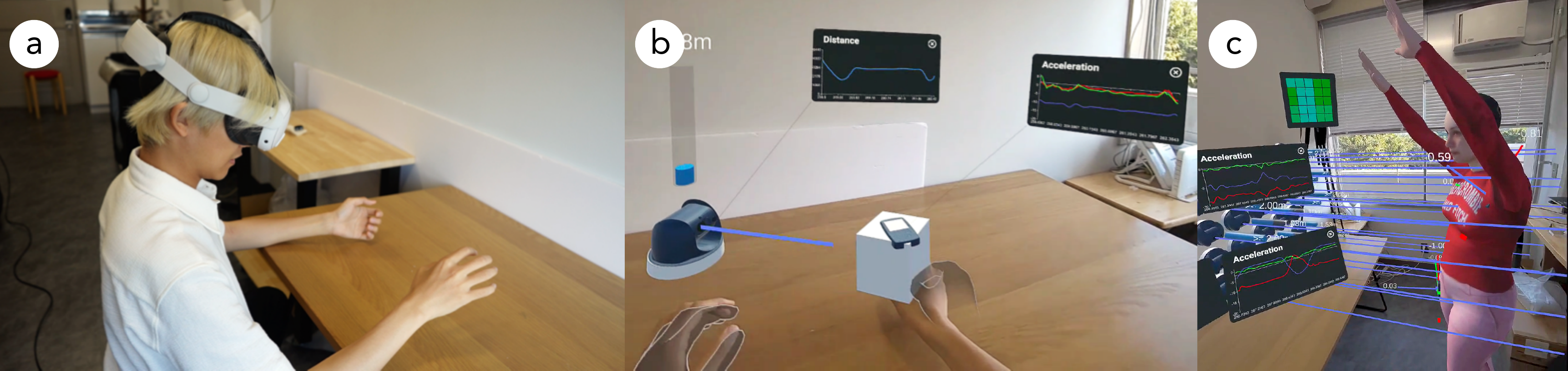}
  \caption{Users (a) can easily inspect sensor data (b) with different placements and orientations through HMDs, and use the models and prefabs to build complicated system experiments (c).}
  \Description{The figure contains a...}
  \label{fig:teaser}
\end{teaserfigure}


\maketitle

\section{Introduction}
Advancements in electronics and machine learning are steering us toward a future filled with trillions of Internet of Things (IoT) devices, furthering the vision of ubiquitous computing. At the core of these intelligent systems, devices are now equipped with enhanced HAR capabilities. This progress has drawn the attention of numerous researchers and enthusiasts, who are designing systems that incorporate sensors into various objects and body surfaces. However, designing a reliable and accurate HAR system is not a trivial task, especially for those utilizing non-visual sensors for marginal and specific scenarios, which are often plagued by the insufficient amount of data to train a reliable machine learning model.

Virtual sensing is one of the ways to enlarge the dataset. However, current virtual sensing systems, represented by IMUtube \cite{IMUTube}, are often highly integrated and designed for automated batch processing of data, lacking of a lightweight and easy-to-use interface. On the other hand, few systems that utilize AR technology focus on the data layer \cite{SensorViz}. We aim for a system that lowers the barriers for users in interpreting and pre-processing sensor data, providing a high-level preview of system inputs. XR technology provides an excellent interface for using virtual sensors for simplified data collection. Users can not only preview sensor placements around them through pass-through but also collect data easily without any wiring or interconnected components. The flexibility of the virtual space allows us to easily import human motion animations from motion capture databases into mixed-reality scenarios. These motion animations can provide referable virtual data without the need to conduct user studies, assisting in the early stages of system design.

Our demonstrated system, \textit{Vsens Reality}\footnote{Vsens Reality demo: \url{https://youtu.be/3HCpBNiLSNM}}, utilizes only the real-time meshing and plane detection capabilities of HMDs, allowing users to design their sensor setups in an environment as close to reality as possible. In the future, our ultimate goal is to propose a new development pipeline for HAR system design with more diverse sensor types in the XR and virtual environments, in order to replace or enhance the design of time-consuming and labor-intensive user studies.

\section{Key Characteristics}
\begin{figure*}[htbp]
\centering
  \includegraphics[width=2\columnwidth]{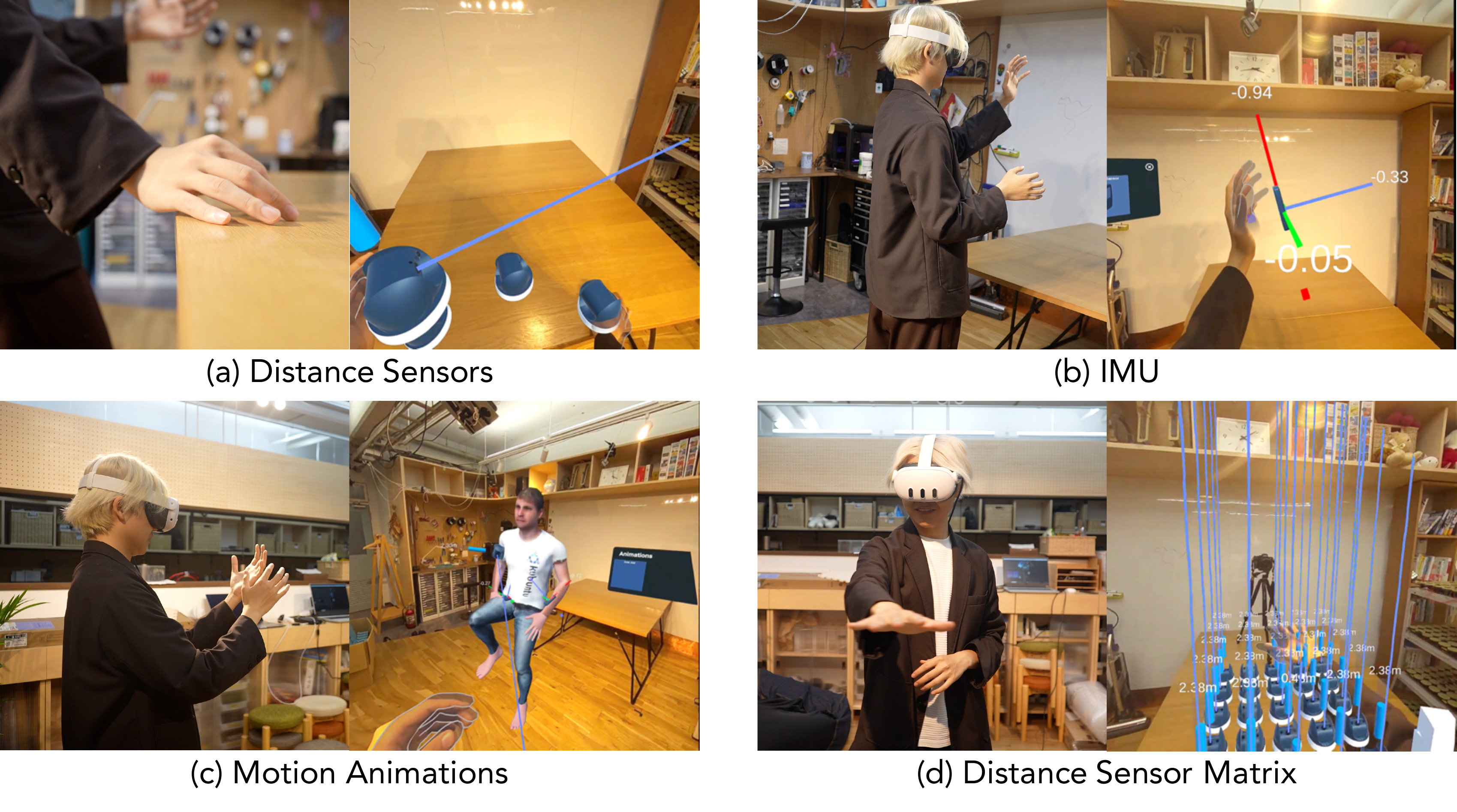}
   \caption{Scenarios of user virtual sensor with the mixed reality.}~\label{fig:demonstrations}
   \Description{The figure contains a...}
\end{figure*}

Our entire system relies solely on an HMD with the pass-through feature as its hardware requirement. All the functionalities can be accessed through gestures without controllers. The system was tested on \textit{Meta Quest 3} and without migration issues between different versions of the platforms. The software part was developed with \textit{Unity3D} with the \textit{Meta XR All-in-One SDK}.

\textbf{Distance Sensor.} We implement a type of distance sensor inside the toolkit. It features a base that can be attached to various surfaces and a rotatable ray-cast emitter. In the virtual space, users can intuitively view the length of the ray and visualize data through a bar graph above the sensor. The combinations of the distance sensors can be saved as various prefabs, such as matrices, for use in designing the HAR systems.

\textbf{Inertial Measurement Unit (IMU).} IMUs are a frequently studied type of virtual sensor. By smoothing the positional data of the sensor using interpolation and then performing a second differentiation, we obtain the local acceleration of the virtual IMU. The three axes around the IMU are marked with red, green, and blue, with the magnitude of the acceleration labeled at the top of each axis. Additionally, we have placed three small movable cubes along each axis to represent the acceleration, providing further enhanced visualization.

\textbf{Animations \& Models.} In the demonstration application, we have included a variety of animations of motion exercises and human models of different body shapes. Users can easily observe the data changes from different types of sensors applied to them. Additionally, we have added small objects with physical simulations to serve as IMU attachment tests.

\textbf{Live Data Visualization.} 
Each sensor option includes a panel that visualizes the sensor signal in the form of dynamic curves. These panels can display the trends and time-related features of the data over a period, similar to how traditional physical sensors are shown in a serial plotter.

\section{Future Vision}
The proposed \textit{Vsens Reality} is merely a prototype concept that introduces virtual sensing technology into XR interactions in the current stage. More development is needed to enhance its usability, and extensive evaluation is required to verify whether the generated virtual data is sufficiently realistic. Additionally, we need more validation, such as workshops, to understand which features are truly needed by developers with different levels of knowledge, and to determine whether this design toolkit can genuinely assist in the design process.


\bibliographystyle{ACM-Reference-Format}
\bibliography{sample-base}
\end{document}